\documentclass[prl,showpacs,twocolumn]{revtex4-1}
\usepackage{amsmath,graphicx}

\def \dif {\mathrm{d}}

\begin{document}

\title{Super-Rough Glassy Phase of the Random Field XY Model in Two Dimensions}

\author{ Anthony Perret$^1$, Zoran Ristivojevic$^2$, Pierre Le Doussal$^2$, Gr\'egory Schehr$^1$, and Kay J. Wiese$^2$}

\affiliation{$^1$Laboratoire de Physique Th\'{e}orique et Mod{\`e}les Statistiques, CNRS-Universit\'{e} Paris-Sud, B{\^at.}~100, 91405 Orsay France}
\affiliation{$^2$Laboratoire de Physique Th\'{e}orique--CNRS, Ecole Normale Sup\'{e}rieure, 24 rue Lhomond, 75005 Paris, France}

\begin{abstract}
We study both analytically, using the renormalization group (RG) to two loop order, and numerically, using an exact polynomial algorithm, the disorder-induced glass phase of the two-dimensional XY model with quenched random symmetry-breaking fields and without vortices. In the super-rough glassy phase, i.e.~below the critical temperature $T_c$, the disorder and thermally averaged correlation function $B(r)$ of the phase field $\theta({\bf x})$, $B(r) = \overline{\langle [ \theta({\bf x}) - \theta({\bf x}+ {\bf r}) ]^2\rangle}$ behaves, for $r \gg a$, as $B(r) \simeq A(\tau) \, \ln^2 (r/a)$ where $r = |{\bf r}|$ and $a$ is a microscopic length scale. We derive the RG equations up to
cubic order in $\tau = (T_c-T)/T_c$ and predict
the universal amplitude ${A}(\tau) = 2\tau^2-2\tau^3 + {\cal O}(\tau^4)$. The universality of $A(\tau)$ results from nontrivial cancellations between nonuniversal constants of RG equations. Using an exact polynomial algorithm on an equivalent dimer version of the model we compute ${A}(\tau)$ numerically and obtain a remarkable agreement with our analytical prediction, up to $\tau \approx 0.5$.
\end{abstract}
\maketitle

Disordered elastic systems are relevant to describe various experimental situations ranging, for interfaces, from domain walls in ferromagnetic \cite{Lemerle98} or ferroelectric \cite{Paruch05} systems, contact lines in wetting \cite{Moulinet05} to propagating cracks \cite{Santucci07} and, for periodic structures, from vortex lattices (VLs) in type-II superconductors \cite{Blatter94} and Wigner crystals \cite{GiamarchiWigner04}, to charge or spin density waves \cite{Gruner88}. In most of these systems, the large scale properties are described by a zero temperature fixed point, which can be described analytically using the functional renormalization group~\cite{LeDoussalReviewFRG}. This latter has led to very accurate predictions, e.g. concerning various exponents, which could be, in some cases, successfully confronted to experiments or numerical simulations~\cite{WieseLeDoussal07}.

In some cases, however, thermal fluctuations play an important role: this is the case for systems where the exponent describing
the scale dependence of the free energy fluctuations $\Delta F \sim L^\theta$ is $\theta =0$. It is then crucial to study the interplay between disorder and thermal fluctuations.
\begin{figure}[h]
\includegraphics[width=0.8\linewidth]{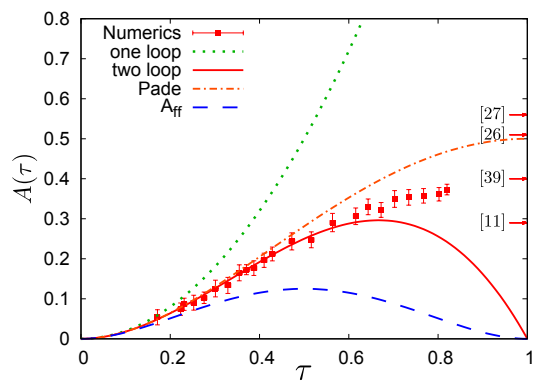}
\caption{The amplitude ${A}(\tau)$, characterizing the super-rough phase. The squares indicate the numerical estimates obtained here using an exact polynomial algorithm~\cite{Propp}. The 'one loop' curve indicates the one loop result $A(\tau)=2 \tau^2$ while the 'two loop' curve shows the two loop result (\ref{amplitude}) obtained here (we also show a Pad{\'e} resummation of it). $A_{\rm ff}$ is the result obtained in Ref.~\cite{LeDoussal+07} from translating to (\ref{H}) the free fermion calculation of Ref.~\cite{Guruswamy+00}. We also show the values obtained at $T=0$ in the corresponding references.}\label{fig_amplitude}
\end{figure}
While at zero temperature, Monte Carlo (MC) simulations, which are hampered by extremely long equilibration times, could be circumvented by the use of powerful algorithms to compute directly the ground states using combinatorial optimization, the latter are of little use to study finite temperature properties. Here we consider a prototype of such situations,
the classical 2D XY model with quenched random fields, known as the Cardy-Ostlund (CO) model  \cite{Cardy+82}. It describes a wide class of systems including 2D periodic disordered elastic systems, such as a randomly pinned planar array of vortex lines \cite{Hwa+94,Bolle+99Nature}, surfaces of crystals with quenched disorder \cite{Toner+90}, random bond dimer models~\cite{Bogner2004} and noninteracting disordered fermions in 2D~\cite{Guruswamy+00,LeDoussal+07}. In terms of a real phase field $\theta(\mathbf{x}) \in (-\infty,\infty)$, the CO model is defined by the partition function $Z = \int {\cal D} \theta \, e^{-H/T}$ with the Hamiltonian
\begin{align}\label{H}
H=\int\dif^2 x\left[\frac{\kappa}{2}\left(\nabla\theta\right)^2-\mathbf{f}\cdot\nabla\theta- \frac{1}{a}\left(\xi{e}^{i\theta}+\mathrm{H.c.}\right)\right] \;.
\end{align}
Here $\kappa$ is the elastic constant, $a$ is the short-length-scale cutoff, and $\mathbf{f}$ and $\xi$ are quenched Gaussian random fields. Their nonzero correlations are given by
\begin{align}
&\overline{f^i(\mathbf{x})f^j(\mathbf{y})}= T^2 \frac{\sigma}{2 \pi} \delta^{ij}\delta(\mathbf{x}-\mathbf{y}) \;, \\
&\overline{\xi(\mathbf{x})\xi^*(\mathbf{y})}=T^2 \frac{g}{2 \pi} \delta(\mathbf{x-y}) \;,
\end{align}
where $i,j\in\{1,2\}$ denote the components of $\mathbf{f}$, $T$ is the temperature, and $\overline{.\phantom{1}.\phantom{1}.}$ denotes the disorder average. The disorder $\mathbf{f}$ must be introduced in the model as it is generated by the symmetry-breaking field under coarse graining \cite{Cardy+82}.
The CO model exhibits a transition at a critical temperature $T_c = 4 \pi \kappa$ between a high-temperature phase, where disorder is irrelevant, and a low-temperature disorder induced glass phase. It is described by a line of fixed points indexed by $T$, which, thanks to the statistical tilt symmetry (STS), is not renormalized ($\theta = 0$ to all orders in perturbation theory). It displays many features of glassy systems, e.g., universal susceptibility fluctuations \cite{ZengLeathHwa1999}, and aging~\cite{SchehrLeDoussal2004,SchehrRieger2005}.

The most striking effect of disorder on the statics concerns the two-point correlation function (CF) $B(r) = \overline{\langle [ \theta({\bf x}) - \theta({\bf x}+ {\bf r}) ]^2\rangle}$. While for $T > T_c$ the interface is logarithmically rough, $B(r) \approx {4T}/{T_c} \ln{(r/a)}$, it becomes super rough for $T~<~T_c$ where
\begin{eqnarray}\label{def_log_sq}
B(r) = A(\tau) \, \ln^2 (r/a) + {\cal O}[\ln(r/a)] \;,
\end{eqnarray}
with $\tau = (T_c-T)/T_c$. The temperature $T$ is determined by the connected CF $B_c(r) = \overline{\langle [ \theta({\bf x}) - \theta({\bf x}+ {\bf r}) ]^2\rangle_c} \simeq (4T/T_c) \ln{(r/a)}$. A physical realization of (\ref{H}) is the VL confined in a superconducting film with a parallel magnetic field \cite{mpa_fisher,Hwa+94,GiamarchiLeDoussal95}. Such a geometry was realized experimentally on mesoscopic devices \cite{Bolle+99Nature}. The $\ln^2(r)$ growth in ~(\ref{def_log_sq}) results in a loss of translational order. Indeed it can be shown~\cite{Toner+90,Ristivojevic+12unpub} that (\ref{def_log_sq}) implies that the CF of the order parameter for translational
order of the VL decays faster than any power law $\ln\left(\overline{\langle e^{i q(\theta({\bf x +r})-\theta({\bf x}))}\rangle}\right) \sim \ln^2(r)$. This can be probed by neutron scattering experiments or direct observation via STM or scanning superconducting quantum interference device probes~\cite{finkler+12}.

Although the amplitude $A(\tau)$ of this intriguing $\ln^2(r)$ has been the subject of numerous studies \cite{carpentier+97,Zeng+96,Rieger+97,ZengLeathHwa1999,Lancaster+07}, none of them was able to establish a quantitative comparison between analytical results and numerical simulations, and there are several reasons for this gap. The first one is that $A(\tau)$ was, analytically, only known, at lowest order in $\tau$, $A(\tau) = 2 \tau^2 +{\cal O}(\tau^3)$ \footnote{note that various values of the $\tau^2$ coefficient have appeared in the literature before it was fixed in \cite{carpentier+97}}: its domain of validity is thus restricted to a narrow region close to $T_c$, where the amplitude of the $\ln^2(r)$ term is small and thus hard to isolate accurately from the subleading logarithmic correction in (\ref{def_log_sq}). The second reason is that numerics is very delicate, given that standard MC simulations are quite inefficient for $T < T_c$. Fortunately, there exists an exact polynomial algorithm, called the domino shuffling algorithm (DSA), which allows us to sample directly the related random bond dimer model, without running MC simulations. This algorithm was used in Ref.~\cite{ZengLeathHwa1999}, which showed that $A(\tau) \propto \tau^2$, without estimating the prefactor. Notice that, in its original formulation as used in Ref.~\cite{ZengLeathHwa1999}, the DSA suffers from strong finite size effects reminiscent of the arctic circle phenomenon~\cite{Jockush} in the pure dimer model.

In this Letter, we perform a quantitative comparison, in a wide temperature range, between analytical and numerical predictions. Such a comparison is rendered possible (i) thanks to a precise calculation of $A(\tau)$, using various RG schemes, yielding the following expression to two loop order:
\begin{align}\label{amplitude}
A(\tau)=2\tau^2-2\tau^3+\mathcal{O}(\tau^4) \;,
\end{align}
and (ii) thanks to a careful Fourier analysis of the two-point CF. It is computed here using an improvement of the DSA~\cite{janvresse}, where finite size effects are significantly reduced. The result of this comparison is shown in Fig.~\ref{fig_amplitude}. We see a remarkable agreement between both approaches, even far beyond $T_c$ down to $\tau \approx 0.5$.

Our analytical study determines the RG equations for the model (\ref{H}) to two-loop order. We use the replica method to treat the disorder \cite{Giamarchi} and obtain the replicated Hamiltonian $H^{rep}=H_0^{rep}+H_1^{rep}$,
with the harmonic part
\begin{align}\label{H0}
\frac{H_0^{rep}}{T}=&\sum_{\alpha\beta}\int\dif^2x\Big\{ \frac{\kappa}{2T}\delta_{\alpha\beta}\left[ (\nabla\theta_\alpha({\bf x}))^2+m^2(\theta_\alpha({\bf x}))^2\right]\notag\\
&-\frac{\sigma}{4 \pi}\nabla\theta_\alpha({\bf x}) \cdot\nabla\theta_\beta({\bf x}) \Big\}.
\end{align}
The mass $m$ is introduced as an infrared cutoff while $\alpha$ and $\beta$ denote $n$ replica indices. Both $m$ and $n$ are set to zero at the end. The anharmonic part reads
\begin{align}
\label{H1}
\frac{H_1^{rep}}{T}=-\frac{g}{2 \pi a^2}\sum_{\alpha\beta} \int\dif^2x\cos(\theta_\alpha({\bf x})-\theta_\beta({\bf x})).
\end{align}
We compute the two CFs:
\begin{align} \label{calG}
&{\cal G}(\mathbf{x}) = \overline{\langle \theta(\mathbf{x}) \theta(0) \rangle_H } - \overline{ \langle \theta(\mathbf{x}) \rangle_H \langle \theta(0) \rangle_H}, \\
&{\cal G}_0(\mathbf{x}) = \overline{\langle \theta(\mathbf{x}) \rangle_H \langle \theta(0) \rangle_H },
\end{align}
where the former measures the (disorder averaged) thermal fluctuations while the latter measures the fluctuations due to disorder of the (thermally averaged) phase field. These CFs can be obtained from CFs of replicated fields
by decomposing $
{\cal G}_{\alpha\beta}(\mathbf{x}):=\langle \langle \theta_\alpha(\mathbf{x})\theta_\beta(0)\rangle\rangle= \delta_{\alpha\beta}{\cal G}(\mathbf{x})+{\cal G}_0(\mathbf{x})$, where
${\cal G}(\mathbf{x})$ is called the connected part and
${\cal G}_0(\mathbf{x})$ the off-diagonal part. To compute them we use the harmonic part $H_0^{rep}$ as the "free" theory and treat $H_1^{rep}$ in perturbation theory in $g$. Here we denote by $\langle  \langle .. \rangle \rangle$ averages over the complete Hamiltonian $H^{rep}$ and by $\langle .. \rangle$ averages over the free part $H_0^{rep}$.

We start by computing the CF for $g=0$ and find
\begin{align}\label{G-fourier}
G_{\alpha\beta}(\mathbf{q})= \frac{T}{\kappa}\frac{\delta_{\alpha\beta}} {q^2+m^2}+\frac{\sigma T^2}{2\pi\kappa^2} \frac{q^2}{(q^2+m^2)^2}+\mathcal{O}(n).
\end{align}
In real space one obtains $G_{\alpha\beta}(\mathbf{x})=\langle \theta_\alpha(\mathbf{x})\theta_\beta(0)\rangle= \delta_{\alpha\beta}G(\mathbf{x}) +G_0(\mathbf{x})$.
The connected part behaves at small distances $|\mathbf{x}|\ll (c \, m)^{-1}$ as
\begin{align}\label{G-smallx}
G(\mathbf{x})=-(1-\tau)\ln \left[c^2 m^2(x^2+a^2)\right ],
\end{align}
with $c=e^{\gamma_E}/2$ and $\gamma_E$ is the Euler constant. In (\ref{G-smallx}) we have introduced the ultraviolet regularization by the parameter $a$ \cite{Amit+80}. The off-diagonal part of the CF at small distances reads
$
G_0(\mathbf{x})=-2\sigma(1-\tau)^2 \ln\left[ec^2m^2({x}^2+a^2)\right]+\mathcal{O}(n). \label{G0x}
$

The STS of the model manifests itself by the invariance of the non-linear part $H_1^{rep}$ under the change $\theta_\alpha(\mathbf{x}) \to \theta_\alpha(\mathbf{x}) + \phi(\mathbf{x})$ for an arbitrary $\phi(\mathbf{x})$. As discussed in~\cite{Schulz+88,Hwa+94,carpentier+97} it implies two properties: (i) $G_0(\mathbf{x})$ does not appear to any order in perturbation theory in $g$ and (ii) the disorder-averaged thermal CF is ${\cal G}(\mathbf{x}) = G(\mathbf{x})$ to all orders in $g$. This implies that $T$ can be {\it measured} from the amplitude of the logarithm in ${\cal G}(\mathbf{x}) \simeq 2 (1-\tau) \ln x$ at large $x$. Because of property (i) $G_0(\mathbf{x})$ only receives additive corrections, e.g.~corrections to $\sigma$ which, in the present model, change its above logarithmic behavior into a squared-logarithm behavior for ${\cal G}_0(\mathbf{x})$, obtained below.

To obtain the scaling equations beyond lowest order we compute the effective action up to $\mathcal{O}(g^3)$, which reads
\begin{align}\label{Gamma-finaldiv}
\Gamma=&\sum_{\alpha\beta}\int\dif^2x\Big\{ \frac{\kappa}{2T}\delta_{\alpha\beta}\left[ (\nabla\theta_\alpha)^2+m^2(\theta_\alpha)^2\right]\notag\\
&-\frac{\sigma_R}{4 \pi}\nabla\theta_\alpha \cdot\nabla\theta_\beta-\frac{g_R}{2 \pi} c^2m^2\cos (\theta_\alpha-\theta_\beta)\Big\}
\end{align}
in terms of \textit{renormalized} couplings $g_R$ and $\sigma_R$. Their explicit dependence on the bare parameters leads to the following scaling equations in terms of the scale
$\ell=-\ln m$:
\begin{align}
\label{RGgfinal}
&\frac{\dif g_R}{\dif\ell}=2\tau g_R - A g_R^2-B\tau g_R ^2+ C g_R^3,\\
\label{RGsigmafinal}
&\frac{\dif\sigma_R}{\dif\ell}=D g_R^2+E \tau g_R^2-F g_R^3,
\end{align}
and ${\dif \tau}/{\dif\ell}=0$, which encodes the exact result $\mathcal{G}(x)=G(x)$ from STS. Remarkably, although $A, B, \ldots, F$ are \textit{nonuniversal} constants, they satisfy the \textit{universal} ratios
\begin{align}\label{inv}
A^2/D=8,\quad A^2/C=4,\quad F + BD -  AE/2 = 0 \;,
\end{align}
which will ultimately enter into the expression of physical quantities, including $A(\tau)$ in (\ref{def_log_sq}). We have obtained these values through three different regularization schemes, for details see Ref.~\cite{Ristivojevic+12}. These equations generalize to two-loop order the one-loop equations obtained in~\cite{Cardy+82,Toner+90,Hwa+94,carpentier+97,SchehrLeDoussal2003}. From (\ref{RGgfinal}) we see that the model has a transition at $\tau=0$, i.e.~$T=T_c$. For $T > T_c$ the renormalized coupling $g_R(\ell)$ flows to zero, while for $T<T_c$ it flows to a finite value $g_R^*=2\tau/A +(4C-2AB) \tau^2/A^3+\mathcal{O}(\tau^3)$. The asymptotic solution of (\ref{RGsigmafinal}) is
$\sigma_R(\ell) \simeq \sigma_0 + [\dif\sigma(g^*_R)/\dif\ell]\ell$: while $\sigma_0$ is nonuniversal and leads only to logarithmic growth of the CF, the second term yields the $\ln^2(r)$ growth~\cite{Toner+90}. To estimate the off-diagonal CF at a given wave vector $q$, one considers the limit $m \ll q$ and argues that $q$ sets the scale $\ell^*= \ln[1/(aq)]$ at which one stops the flow. Replacing $\sigma$ by its effective value at that scale, i.e.~$\sigma \to \sigma_R(\ell)$ we get, from (\ref{G-fourier}) the small $q$ behavior
$
{\cal G}_0(\mathbf{q})\simeq 8 \pi (1-\tau)^2 \frac{\dif\sigma(g^*_R)}{\dif\ell} \frac{\ln[1/(q a)]}{q^2},
$
which leads to (\ref{def_log_sq}) and (\ref{amplitude}). The amplitude (\ref{amplitude}) is \textit{universal} thanks to the remarkable combination of nonuniversal constants in $\dif\sigma(g^*_R)/\dif\ell$. Equation (\ref{def_log_sq}) can be obtained more rigorously by calculating the two-point function~\cite{Ristivojevic+12}.

We have performed simulations to estimate numerically the amplitude $A(\tau)$ and compare it with (\ref{amplitude}). For that purpose, we use the mapping between (\ref{H}) and a weighted dimer model defined on a 2D lattice~\cite{ZengLeathHwa1999,Bogner2004}, for which there exists a polynomial DSA~\cite{Propp}. For technical reasons, it is designed for a special lattice $A_L$ called the Aztec diamond of size $L$ [see Fig.~\ref{fig_diamond}(a)]. To each bond between nearest neighbors $({\bf r}, {\bf r'})$ on $A_L$ we assign a quenched random variable $\epsilon_{{\bf r}, {\bf r'}}$: here we consider independent Gaussian variables of zero mean and unit variance. The dimer model consists of all complete dimer coverings of $A_L$, where the weight $W(\cal C)$ of a dimer covering ${\cal C}$ is given by
\begin{eqnarray}\label{def_weight}
W({\cal C}) = \frac{1}{Z_L(\epsilon)} \exp{(-H_d/T_d)} \;, \; H_d = \sum_{({\bf r}, {\bf r'}) \in {\cal C}} \epsilon_{{\bf r}, {\bf r'}} \;,
\end{eqnarray}
where $Z_L(\epsilon)$ is the partition function. Hence the limit $T_d \to 0$ corresponds to a "strong" disorder regime while the limit $T_d \to \infty$ corresponds to dimer coverings with uniform weights. The DSA generates uncorrelated dimer configurations, directly sampled with the equilibrium weight (\ref{def_weight}), without the need to run a slow MC algorithm. In addition, this is a polynomial algorithm (with a computational time $\sim L^3$). The dimer covering of the Aztec diamond is, however, known to suffer from strong finite size effects~\cite{Jockush}. Here we minimize significantly these effects by using a recent improvement of the DSA which allows for the existence of bonds with zero weight \cite{janvresse}. We use it to study the random dimer model directly on a square lattice, which exhibits less pronounced finite-size effects~\cite{PerretSchehr2011}.
This algorithm is very flexible and will be very useful to study other dimer systems in various 2D geometries.

To a given dimer covering ${\cal C}$, we assign a discrete height field, defined on the center of the squares [see Fig.~\ref{fig_diamond}(a)], i.e.~on the dual lattice $A_L^D$ of $A_L$, as follows~\cite{henley}. The bonds of $A_L^D$ are oriented such that the unit cells of $A_L^D$ that enclose the blue sites of $A_L$ are circled counterclockwise. Assign $-3$ to the difference of neighboring heights along the oriented bonds if a dimer is crossed and $+1$ otherwise. This yields single-valued heights up to an overall constant, the heights on the boundaries of $A_L$ being then fixed as in Fig.~\ref{fig_diamond}(a). This defines a height field $\widetilde{H}_{\bf r} \equiv \widetilde{H}_{ij}$, with ${\bf r} = i {\bf u_x} + j {\bf u_y}$ and the relative height $h = \widetilde{H} - \prec \widetilde{H} \succ$ where $\prec \widetilde{H} \succ$ is a spatial average of $\widetilde{H}$ over $A_L$. For uniform dimer coverings, corresponding to $\epsilon_{{\bf r},{\bf r'}} = 0$ or $T_d \to \infty$, one can show that the fluctuations of $h$ in the continuum limit {(and in the bulk) are described by a Gaussian free field \cite{henley,kenyon_gff}, {i. e.} by the Hamiltonian in~(\ref{H}) without disorder (${\mathbf f} = 0$, $\xi = 0$) at $\tau = 0$. For inhomogeneous random bonds $\epsilon_{{\bf r},{\bf r'}}$ one expects instead that in the continuum limit, the fluctuations of $h$ are described by the CO model (\ref{H}) with the substitution $\theta \to h \times 2 \pi/4$~\cite{ZengLeathHwa1999,Bogner2004}. This factor~$2 \pi/4$ is required because the energy associated to the height configurations~(\ref{def_weight}) is invariant under a global shift $h \to h + 4$.

\begin{figure}[h]
\begin{center}
\includegraphics[width=\linewidth]{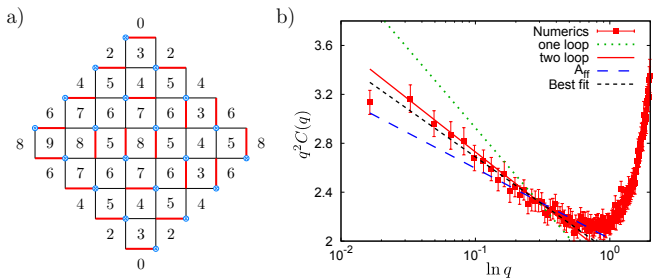}
\caption{a) Dimer covering of an Aztec diamond of size $4$, $A_4$. The blue points allow us to define the height field, which are the integers on the dual lattice $A_4^D$. b) Plot of $q^2 C(q)$ as a function of $\ln q$. The squares correspond to our numerical data obtained for lattice size $L=384$. The slope of the straight line indexed by 'one loop' and 'two loop' is given respectively by the one loop ${A}(\tau) = 2 \tau^2$ and the two loop estimate~(\ref{amplitude}), while $A_{\rm ff}$ corresponds to Ref.~\cite{Guruswamy+00}. Without a $\ln^2$ term in (\ref{def_log_sq}), one would expect a straight line with vanishing slope: this is ruled out by our data and demonstrates the existence of the super-rough phase.}
\label{fig_diamond}
\end{center}
\end{figure}

The temperature $T_d$ of the dimer model does not coincide with the temperature of~(\ref{H}). To compute $\tau$ we use the STS (\ref{G-smallx}) and measure
\begin{eqnarray}
W^2_T = \frac{1}{L^2}\sum_{\bf r} \overline{\langle h_{\bf r}^2 \rangle -  \langle h_{\bf r}\rangle\langle h_{\bf r}\rangle} \simeq  2(1-\tau) \ln{L} \;,
\end{eqnarray}
which provides a precise estimate of $\tau$. We have checked that our numerical estimate is in good agreement with the analytic results for $\tau$ and for other thermodynamical observables obtained in Ref.~\cite{Bogner2004}.

We want to compute numerically the amplitude of the $\ln^2(r)$ term in~(\ref{def_log_sq}). Extracting this amplitude precisely from $B(r)$ is however difficult, since the subleading corrections are of order ${\cal O}(\ln r)$. The calculation is more accurate in Fourier space \cite{Zeng+96,Schwartz}, defining $\hat h_{\bf q} = L^{-2} \sum_{\bf r} h_{\bf r} e^{i {\bf q} \cdot {\bf r}}$. The CF $C(q)$ of these Fourier components is expected, from~(\ref{def_log_sq}),  to behave for small $q$ as
\begin{eqnarray}\label{correl_fourier}
C(q) = \overline{\langle \hat h_{\bf q}\rangle \langle \hat h_{-{\bf q}}\rangle}  \simeq \frac{8}{\pi} {A}(\tau) \frac{\ln{(1/q)}}{q^2} + {\cal O}(q^{-2})\;,
\end{eqnarray}
where $q = |{\bf q}|$. In Fig.~\ref{fig_diamond}(b), we plot of $q^2 C(q)$ as a function  of $\ln{q}$ for $\tau \approx 0.33$ ($T_d = 0.25$). These data have been obtained for a system size $L = 384$ by averaging over $10^5$ realizations of the random bonds $\epsilon_{{\bf r}, {\bf r'}}$'s. They support the expected behavior in (\ref{correl_fourier}) for small $q$, $q \lesssim 1$: they are indeed well described by a straight line, $q^2 C(q) = -8{A}(\tau)/\pi \ln{q} + b_0$. Note that the downwards bending for the smallest $q$'s is a finite size effect. In Fig.~\ref{fig_diamond}(b) we also show four different straight lines corresponding to different couples $[{A}(\tau),b_0]$. The line indexed by 'Best fit' corresponds to the best fit of these data by a straight line: the value of ${A}(\tau)$ obtained in this way corresponds to the squares on Fig.~\ref{fig_amplitude}. In the three other cases, the slope of this straight line is evaluated from the one- and two-loop (\ref{amplitude}) results respectively, while the straight line indexed by '$A_{\rm ff}$' corresponds to the slope computed in \cite{LeDoussal+07} from the result in Ref.~\cite{Guruswamy+00}, with $A_{\rm ff}(\tau) = 2 \tau^2 (1-\tau)^2$.
In all cases the constant $b_0$ is a fitting parameter.
One clearly sees that the two loop result is a significant improvement over the one loop result and describes very well our numerical data. Clearly, $A_{\rm ff}$ underestimates our data.

Let us now discuss the numerical results for ${A}(\tau)$ in Fig.~\ref{fig_amplitude}.
As compared to Ref.~\cite{ZengLeathHwa1999}, here we can discuss a much broader range of values of $\tau$ which extends deep
into the glass phase. First we observe that our two loop result is in very good agreement with our numerics up to $\tau \approx 0.5$. In contrast, the curve $A_{\rm ff}(\tau)$ is significantly smaller than our numerical values and can be ruled out. For smaller temperature, $\tau \gtrsim 0.5$ the discrepancy between (\ref{amplitude}) and the numerical value increases, as expected. In Fig.~\ref{fig_amplitude} we have also quoted the numerical values which were obtained independently at zero temperature by an exact ground state calculation for the solid-on-solid model on a disordered substrate in~\cite{Zeng+96,Rieger+97,Schwartz}. This model is also described, in the continuum limit, by the model~(\ref{H}). In particular, our data match smoothly with the most recent numerical estimate obtained in Ref.~\cite{Schwartz}, yielding ${A}(\tau = 0) = 0.39$. We also show an estimate based on a one-loop functional RG calculation at $T=0$ \cite{LeDoussal+07}. The fact that
the two loop formula (\ref{amplitude}) vanishes at $T=0$ is of course an unphysical feature, which can be cured by considering
various guesses or Pad\'e (i.e.~rational functions in $\tau$) approximations which have the same expansion as (\ref{amplitude}) to order
$\tau^3$. One such formula $A(\tau) = 2\tau^2(1-{\tau}/{2})^2$ is plotted in Fig.~\ref{fig_amplitude}: being simple, it also
has a reasonable structure to correct the result of Ref.~\cite{Guruswamy+00}.

In conclusion, we have obtained an accurate description of the glassy phase of the Cardy-Ostlund model
down to temperatures $T_c/2 \lesssim T$, with excellent agreement for the amplitude of the square logarithm between theory and numerics. Understanding the glass phase below $T_c/2$ is an important challenge for the future.

\acknowledgments{This research was supported by ANR grant 09-BLAN-0097-01/2 and by ANR grant 2011-BS04-013-01 WALKMAT.}

%

\end{document}